\documentclass[9pt,conference]{IEEEtran}
\usepackage{waspaa25}
\usepackage{multirow}
\usepackage[justification=centering]{caption}

\usepackage{bm} % for bold math symbols (incl. Greek letters) with \bm{}

\newcommand{\amy}[1]{\textcolor{black}{#1}}

\title{\amy{Moises-Light: Resource-efficient Band-split U-Net For Music Source Separation}}

\name{Yun-Ning (Amy) Hung$^{1}$,
      Igor Pereira$^{1}$,
      Filip Korzeniowski $^{1}$}
\address{$^{1}$Music AI, USA \;
}

\begin{document}

\maketitle

\begin{abstract}
In recent years, significant advances have been made in music source separation, with model architectures such as dual-path modeling, band-split modules, or transformer layers achieving comparably good results. However, these models often contain a significant number of parameters, posing challenges to devices with limited computational resources in terms of training and practical application. While some lightweight models have been introduced, they generally perform worse compared to their larger counterparts. In this paper, we take inspiration from these recent advances to improve a lightweight model. We demonstrate that with careful design, a lightweight model can achieve comparable SDRs to models with up to 13 times more parameters. Our proposed model, \amy{Moises-Light}, achieves \amy{competitive} results in separating \amy{four} musical stems on the MUSDB-HQ benchmark dataset. The proposed model also demonstrates competitive scalability when using MoisesDB as additional training data.

\end{abstract}

\section{Introduction}
\label{sec:introduction}

Music source separation (MSS) aims to isolate individual sources, such as vocals, drums and bass, from a mixed audio signal. Retrieving individual stems from mixtures of music has numerous practical applications. For example, DJs can remix tracks using separated stems \cite{veire2018raw}, and audio engineers can extract individual sounds for stereo-to-surround upmixing or instrument-wise equalization \cite{rafii2018overview}. Additionally, many \amy{Music Information Retrieval (MIR)} tasks benefit from music source separation, including lyrics transcription \cite{cifka-2024-jam-alt}, pitch estimation \cite{nakano2019joint}, music generation \cite{donahue2023singsong}, vocal melody extraction \cite{wang2024mel}, \amy{and more}.

Recent advancements in music source separation have been driven by innovations in model architecture, particularly through deep learning approaches \cite{hennequin2020spleeter, stoter2019open, takahashi2020d3net, chandna2017monoaural, Jansson2017SingingVS}. For instance, HT Demucs \cite{rouard2022hybrid} processes a hybrid input of waveforms and spectrograms using a model that integrates convolution and transformer modules and has achieved good performance in four-stem source separation. BSRNN \cite{luo2023music} further improved separation quality for \emph{vocals}, \emph{drums}, and \emph{other} stems by using bandsplit modules alongside dual-path RNNs modules for sequence processing at the bottleneck. TFC-TDF UNet v3 \cite{kim2023sound} has set benchmarks using U-Net \cite{ronneberger2015u} architectures that integrate stacks of CNN and linear layers. The current state-of-the-art in source separation is attributed to BS-RoFormer \cite{lu2024music}, which enhances BSRNN by replacing RNN modules with RoPE transformer blocks. BS-RoFormer not only achieves the best results on the MUSDB-HQ benchmark but also scales effectively with additional training data. Furthermore, SCNet \cite{chen2024music}, despite its low model complexity, has achieved results comparable to BS-RoFormer in four-stem separation by incorporating bandsplit techniques with Conformer \cite{conformer} blocks and sequence modeling dual-path RNN modules.

While these models \amy{achieve decent objective scores and separation quality}, they are typically large and costly to train and run. The BS-RoFormer model, in particular, consists of 72.2 million parameters to separate a single stem and requires 16 A100-80G GPUs over four weeks for training \cite{lu2024music}. This not only poses challenges for researchers and institutions with limited computational resources, but also prohibits the deployment of such models on edge devices with limited memory and computational power. 

In contrast, \amy{research in multi-speaker separation} is increasingly focusing on lightweight or resource-efficient approaches. For instance, RE-SepFormer \cite{della2024resource} reduces the computational cost of the state-of-the-art SepRoformer architecture \cite{subakan2021attention}. By operating on non-overlapping blocks and compact latent summaries, this model quarters the number of parameters, while maintaining competitive performance. A similar enhancement was proposed by Yip et al. \cite{yip2024spgm} through the introduction of the Single-Path Global Modulation (SPGM) block. The SPGM blocks halve the number of parameters of the original model, but achieve similar results. Additional ultra-lightweight architectures have also been proposed, such as Fspen \cite{Yang2024FspenAU} and GroupComm-DPRNN \cite{luo2021ultra}. With less than one million parameters, these models are capable of running in real-time and low-resource scenarios.

Music source separation has still seen relatively little research on lightweight or resource-efficient models. Chen et al. \cite{chen2024music} recently introduced DTTNet, a lightweight architecture with only 5 million parameters. This model combines CNN and linear layers from TFC-TDF UNet v3 with the sequence modeling capabilities of BSRNN to largely reduce parameters while maintaining comparable performance compared to the original TFC-TDF UNet v3 and BSRNN. Tong et. al. \cite{tong2024scnet} also tried to shrink the model size to only 10.08 million parameters to train a four-stem separation model. Finally, HS-TasNet \cite{venkatesh2024real} and Wave-U-Net \cite{kim23i_interspeech, StollerED18} are designed to be used in real-time, low-latency scenarios. While efficient, all these models perform inferior to most of the more parameter-heavy models mentioned earlier.

In this paper, we introduce \amy{Moises-Light}, a lightweight architecture that is able to achieve decent source separation results with a small number of parameters. Building on the foundation of DTTNet, we integrate insights from previous research to enhance performance. Specifically, we incorporate band-splitting techniques inspired by BSRNN and BS-RoFormer, integrate RoPE transformer blocks for sequence modeling, adopt the encoder-decoder design from SCNet, and implement training strategies informed by various prior studies. By carefully optimizing architectural design and training strategies, our lightweight model, with around 5 million parameters of a single stem model, achieves \amy{competitive} results on the MUSDB-HQ benchmark dataset. 

In summary, the contributions of this work are threefold:
\begin{itemize}
  \item We enhance the lightweight DTTNet architecture by integrating insights from various source separation research papers.
  \item We introduce an efficient band-splitting module that minimizes \amy{the large amount
of parameters from the original band-splitting} while maintaining effectiveness.
  \item We demonstrate that with these improvements, the final model, \amy{Moises-Light}, achieves \amy{competitive} results on the MUSDB-HQ benchmark, utilizing 13 times fewer parameters than the previous state-of-the-art models, BS-RoFormer, and requiring only half the parameters compared to SCNet.
\end{itemize}

\begin{figure}[t]
  \centering
  \centerline{\includegraphics[width=\columnwidth]{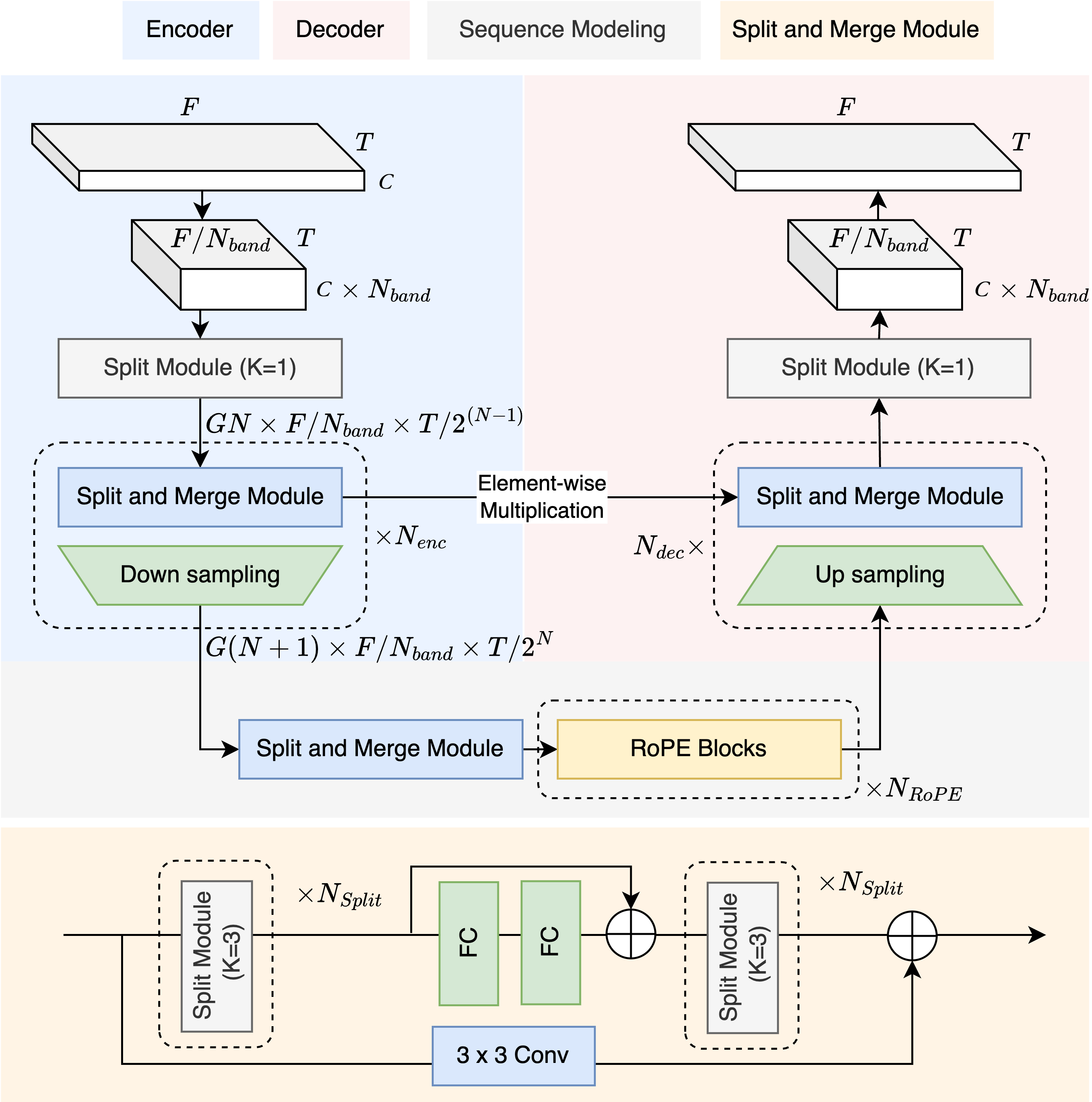}}
  \caption{The overall architecture of our proposed model.}
  \label{fig:moises_light}
\end{figure}

\section{Related Work}
\label{sec:related_works}

\subsection{DTTNet}

DTTNet is a symmetric U-Net architecture consisting of three main components: an encoder, a decoder, and a sequence modeling module. In the encoder, the input complex spectrogram first passes through a $1 \times 1$ convolutional layer to expand the channel features from $C$ into $G$. Following this, $N_{enc}$ encoder layers are employed, each consisting of TFC-TDF V3 block \cite{kim2023sound} and a downsampling block. These layers are designed to extract high-level features, with the TFC (Time-Frequency Convolutions) block capturing local information and TDF (Time-Distributed Fully-connected) block capturing global frequency information. Each encoder layer expands the channel dimension from $G \times N$ to $G \times (N+1)$ while reducing the time and frequency dimension from $T / 2^{N-1}$ to $T / 2^{N}$ and $F / 2^{N-1}$ to $F / 2^{N}$, respectively, with \amy{$N \in \{1, 2, \ldots, N_{\text{enc}}\}$}. The output of the encoder blocks is processed by the sequence modeling module using one TFC-TDF V3 block and $N_{dual}=3$ dual-path RNNs blocks. Then, $N_{dec}$ decoder layers, each consisting of an upsampling block and a TFC-TDF V3 block using skip-connections from the encoder, decode the feature sequence. Finally, a $1 \times 1$ convolutional layer reduces the channels from $G$ to $C$, directly generating the separated target spectrogram $\hat{Y}\in\mathbb{R}^{C \times F \times T}$. 

\subsection{BS-RoFormer}

BS-RoFormer \cite{lu2024music} features two critical components: band-split/multi-band mask estimation modules and RoPE transformer blocks. The band-split module divides complex spectrograms into $N_{band}$ uneven, non-overlapping subbands, with each subband processed by separate multi-layer perceptrons (MLPs) to extract higher-level features. In contrast, the multi-band mask estimation modules reverse this process by decoding the subbands features back into complex spectrograms. As highlighted in the original paper, this approach aims to enhance the model's ability to refine learned representations across different frequency bands, thereby improving robustness against cross-band vagueness. RoPE transformer blocks, placed between the encoder and decoder, use rotary embeddings and dual-path transformers to model features in both time and frequency dimensions.

\subsection{SCNet}

SCNet \cite{chen2024music} has achieved comparable results in four-stem music source separation while utilizing seven times fewer parameters than BS-RoFormer. The framework of SCNet draws similarities to DTTNet, comprising an encoder, a decoder, and dual-path sequence modeling within the bottleneck. However, unlike BS-RoFormer, which splits audio into 62 subbands, SCNet uses only three subbands: low, mid, and high frequency bands. A sparsity-based encoder, which includes downsampling layers and convolution modules, models each frequency band separately. At the end of each encoder block, a global convolution layer integrates the information from each subband. In each decoder block, an upsampling layer combined with a fusion skip connection gradually decodes the information back to the target spectrogram. This approach significantly enhances separation performance while reducing computational demands.

\begin{table}[]
    \centering
    \begin{tabular}{c|c|c|c|c|c}
         $N_{band}$ & $N_{enc}$ & $N_{dec}$ & $N_{RoPE}$ & $N_{split\_enc}$ & $N_{split\_dec}$ \\
    \toprule
         4 & 3 & 1 & 5 & 3 & 1
    \end{tabular}
    \caption{Hyperparameters of the proposed model.}
    \label{tab:params}
\end{table}

\section{Model Architecture} 
\label{sec:methods}

Our model architecture is based on DTTNet. We improve the model in three key aspects: first, we enhance the network architecture by incorporating RoPE transformer blocks and multi-band mask estimation; second, we refine the encoder and decoder design by leveraging insights from SCNet \cite{tong2024scnet}; third, we improve the training pipeline by incorporating data augmentation techniques and employing a multi-resolution complex spectrogram mean absolute error (MAE) \cite{guso2022loss} loss function. Figure \ref{fig:moises_light} illustrates the overall architecture of our proposed model, while detailed parameters are provided in Table \ref{tab:params}. In the following sections, we will introduce the design intuition and modifications made at each key stage, step-by-step.

\subsection{Modification Inspired by BS-RoFormer}

RoPE transformer blocks have shown superior performance compared to RNN. Integrating RoPE transformer blocks into sequence modeling is straightforward. By replacing the dual-path RNN blocks within DTTNet with RoPE transformers, we can achieve this integration without significantly increasing the number of parameters.

Although the original band-splitting techniques in BSRNN and BS-RoFormer show promising results, they require a large amount of parameters. In this work, we propose an efficient band-splitting technique to minimize parameter increase. As illustrated in Fig \ref{fig:moises_light}, our approach involves initially dividing the complex spectrogram into $N_{band}$ equal subbands, which are then concatenated across the channel dimension. Split modules, which is a group convolution layer with kernel size $K$ and $N_{band}$ channel groups, are used to replace the first and last convolution layers in DTTNet. Each group independently processes one subband. This method not only optimizes parameter efficiency, but also allows us to remove frequency pooling or upsampling across all DTTNet layers. Moreover, by reducing the frequency bands from the beginning, the frequency fully-connected layer (FC) within the TDF layer now requires $N_{band}$ times fewer parameters. This change also permits an increase in the number of $G$. 

We set $N_{band}$ to $4$ by default and increase $G$ from $32$ to $48$. Additionally, we set $N_{RoPE}$ to 5 by default. We refer to the original encoder-decoder from DTTNet paper as Enc-Dec V1 and our modified version as Enc-Dec V2.

\begin{table*}[t]
\centering
\caption{Ablation study on step-by-step improvement mentioned in Sec. \ref{sec:methods} using MUSDB-HQ dataset.}
\label{tab:ablation}
\sisetup{
    reset-text-series = false, 
    text-series-to-math = true, 
    mode=text,
    tight-spacing=true,
    round-mode=places,
    round-precision=2,
    table-format=2.2,
    table-number-alignment=center
}
\begin{tabular}{lcccccccccccc}
     & Dual-Path & Enc-Dec & Length & Add. Aug & Loss & Vocals & Drums & Bass & Other & Avg \\
     \midrule
    DTTNet & RNN & V1 & 6s & x & L1 & 10.12 & 7.74 & 7.45 & 6.92 & 8.06 \\
    Repro. & RNN & V1 & 6s & x & L1 & 10.01 & 7.51 & 7.64 & 6.83 & 8.00 \\
    \midrule
    \multirow{2}{*}{Sec. 3.1} & RoPE & V1 & 6s & x & L1 & 10.21 & 7.62 & 8.14 & 7.38 & 8.34 \\
                            & RoPE & V2 & 6s & x & L1 & 10.46 & 8.12 & 8.83 & 7.65 & 8.77 \\
    \midrule
    \multirow{2}{*}{Sec. 3.2} & RoPE & V3 & 6s & x & L1 & 10.51 & 9.40 & 9.51 & 7.56 & 9.25 \\
                            & RoPE & V3 & 9s & x & L1 & 10.55 & 9.57 & \textbf{10.11} & 7.58 & 9.49 \\
    \midrule
    \multirow{2}{*}{Sec. 3.3} & RoPE & V2 & 6s & $\checkmark$ & L1 & 10.61 & 9.17 & 9.37 & 7.89 & 9.26 \\
                            & RoPE & V2 & 6s & x & L1+Multi & 10.79 & 8.70 & 9.11 & 7.88 & 9.12 \\
    \midrule
    Proposed       & RoPE & V3 & 9s & $\checkmark$ & L1+Multi & \textbf{10.92} & \textbf{10.93} & 10.04 & \textbf{7.95} & 9.96 \\
    
 \end{tabular}
\end{table*}

\begin{table}[]
\caption{Ablation study on different numbers of splits.}
\label{tab:ablation_split}
\resizebox{0.49\textwidth}{!}{
    \centering
        \begin{tabular}{cccccccccc}
         $N_{band}$ & G & $N_{RoPE}$ & Vocals & Drums & Bass & Other & Avg \\
         \midrule
        2 & 40 & 7 & \textbf{10.63} & 8.10 & 8.47 & 7.85 & 8.76\\
        4 & 48 & 5 & 10.46 & \textbf{8.12} & \textbf{8.83} & 7.65 & \textbf{8.77} \\
        8 & 48 & 6 & 9.83 & 7.72 & 8.62 & \textbf{7.92} & 8.52 \\
        \end{tabular}
}
\centering
\end{table}

\subsection{Modification Inspired by SCNet}

We adopt two key SCNet design ideas to enhance DTTNet. First, SCNet applies band-splitting/merging layers throughout all encoder/decoder layers, rather than limiting them to just the initial encoder and final decoder layers. Second, SCNet employs an asymmetrical architecture, where the encoder is heavier than the decoder, enabling more effective feature extraction from the input.

Inspired by SCNet, we further modify DTTNet by replacing its TFC-TDF V3 blocks with our proposed split and merge modules. Within these modules, all TDF convolutional layers, except those used for skip connections, from the original TFC-TDF V3 blocks are replaced with our split module, featuring $N_{band}$ channel groups to process each subband independently. We also decrease the number of split module layers in the decoder by three times and increase the number of $G$ to have a heavier encoder. That is, $N_{split}$ is set to 3 in encoder while $N_{split}$ is set to 1 in decoder, and $G$ can increase from $48$ to $56$. We also increased the kernel size of the first and last split module layers from $K=1$ to $K=3$ to better capture relationship between frequency bins. We call this modification of encoder and decoder as Enc-Dec V3.

Furthermore, in the original SCNet setting, a longer input length is used. The input length is set to 11 seconds, nearly twice as long as the 5.94 seconds used in DTTNet. Considering the time dimension pooling and dual-path transformer sequence modeling, we hypothesize that a longer input sequence can better capture temporal information. Therefore, we increase the input length from approximately 6 seconds to 9 seconds.

\subsection{Data and Loss function}

While data augmentation has proven effective in various MIR tasks, its application in source separation has primarily been limited to random mixing and random gain adjustment \cite{lu2024music, Dfossez2019DemucsDE, tong2024scnet}. In this work, we incorporate additional techniques, including polarity inversion, pitch shifting, temporal shifting, and channel flipping, implemented by torch\_audiomentations \footnote{\href{https://github.com/asteroid-team/torch-audiomentations/}{https://github.com/asteroid-team/torch-audiomentations}}.

Regarding loss functions, previous studies in music source separation \cite{lu2024music}, speech denoising/dereverberation \cite{SuJF20-0}, and speech generation \cite{yamamoto2020parallel} have demonstrated that frequency domain loss functions significantly improve audio quality. Given that the proposed model operates in the STFT domain, multi-resolution STFT techniques, as discussed in \cite{yamamoto2020parallel}, can prevent overfitting to a fixed STFT setting, facilitate the training process, and result in more natural waveforms using fewer model parameters. Inspired by these findings, we \amy{added} the multi-resolution complex spectrogram mean absolute error (MAE) \cite{guso2022loss} loss function, as employed in BS-RoFormer, \amy{with the L1 loss from the original DTTNet without any weight adjusting}.

\section{Experimental Design} \label{sec:experiment}

To validate our proposed algorithms, we selected the MUSDB-HQ dataset \cite{musdb18-hq}, which has been widely used for benchmarking in the literature. This dataset comprises 150 full-length stereo tracks with a predefined split: 86 tracks for training, 14 for validation, and 50 for evaluation.  In addition to MUSDB-HQ, we expanded our training data using MoisesDB \cite{moisesdb}. We consolidated each track in MoisesDB into four stems (\emph{vocals}, \emph{bass}, \emph{drums}, and \emph{other}) to match the stems in MUSDB-HQ. Both datasets are stereo and sampled at 44.1 kHz.

To create training samples, we segmented each song into chunks of T seconds with $75\%$ overlap, with T set to either 6 or 9. During each epoch, the model iterates through all these samples once. For the Short-Time Fourier Transform (STFT), we used a window size of 6144 samples and a hop length of 1024, truncating the frequency bins to 2048, following the method outlined in \cite{kim2023sound}. During evaluation, we \amy{chunk the audio into T seconds and} utilized a 50\% overlap-add method to ensure audio reconstruction and continuity \cite{lu2024music}.

As evaluation metric, we used Chunk-level SDR (cSDR) \amy{in dB} \cite{stoter20182018}. cSDR computes the Signal-to-Distortion Ratio on 1-second chunks and reports the median across the median SDR over all 1-second chunks in each song.

In terms of training, we employed the AdamW optimizer with a learning rate of $\num{2e-4}$ for all experiments. To further optimize training, we multiply the learning rate by 0.9 if no improvement in validation loss has been observed for 20 epochs. We also incorporated early stopping \amy{if no improvement in validation loss for 50 epochs}. The model with the best cSDR on the validation set was selected. Our experiments were conducted using two H100-80G GPUs with batch sizes of 8 each. The maximum training time for our best model was approximately 5 days for the MUSDB-HQ benchmark. %\amy{As a reference}, BS-RoFormer required a 4-week training period on 16 Nvidia A100-80GB GPUs.

\section{Results} \label{sec:results}

\subsection{Ablation Study} \label{sec:ablation}

We first study the impact of our modifications mentioned in Sec. \ref{sec:methods} using MUSDB-HQ dataset. The results are shown in Table \ref{tab:ablation}. Our reproduced DTTNet model (Repro.) has similar performance compared to the original DTTNet, as shown in the first two rows of Table \ref{tab:ablation}. We further evaluate improvements made by incorporating RoPE blocks and our proposed splitting modules. As indicated in Section 3.1 of Table \ref{tab:ablation}, these modifications contribute an average improvement of 0.7 dB, with the proposed splitting modules significantly boosting performance on \emph{drums}. 

Additionally, we conducted an ablation study to assess the impact of varying the number of splits, $N_{band}$, on performance. Modifying the best model from Sec. 3.1 in Table \ref{tab:ablation}, Table \ref{tab:ablation_split} presents the results for different split values, $N_{band}$. Changing the number of splits affects the model's parameters. To counter this, we reduced or increased $G$ and $N_{RoPE}$ to ensure the total parameter count remained comparable, around 5 million parameters, as depicted in Table \ref{tab:ablation_split}. Vocals separation has better performance when $N_{band}$ is smaller, while Drums and Bass have better performance when $N_{band}=4$. The average scores between $N_{band}=4$ and $N_{band}=2$ is similar. However, since $N_{band}=4$ allows for faster training and inference, we use $N_{band}=4$ for the rest of the experiments in Table \ref{tab:ablation}.

Section 3.2 from Table \ref{tab:ablation} highlights the architecture-wise improvement inspired by SCNet. By integrating the proposed split modules throughout all encoder and decoder layers and adopting an asymmetric encoder/decoder design, we achieve an additional average performance gain of 0.48 dB, with notable improvements in the separation of \emph{drums} and \emph{bass}. Moreover, extending the input sequence enhances the dual-path module's capability to capture inter-frame information, resulting in further performance enhancements, especially for \emph{bass}, which usually has longer note sustain.

Section 3.3 from Table \ref{tab:ablation} demonstrates the impact of adding additional data augmentation techniques and multi-spectrogram loss, building upon the best model from Section 3.1. Additional data augmentation significantly benefits \emph{drums}, resulting in an improvement of nearly 1 dB. During our experiments reproducing DTTNet, we observed that \emph{drums} and \emph{bass} are prone to be trapped in local optima, leading to earlier stopping of learning. By employing additional data augmentation, we can extend the model's learning process. The multi-resolution loss particularly benefits \emph{drums}. Unlike other instruments, drums produce sounds over a broad frequency spectrum with sharper transients. Multi-resolution STFT helps in capturing these details in drum sounds, contributing to a more accurate separation.

\begin{table}[]
\caption{Compared our model with previous literatures on MUSDB-HQ benchmark dataset.}
\label{tab:benchmark}

    \centering
    \resizebox{0.49\textwidth}{!}{
        \begin{tabular}{lccccccc}
         & Vocals & Drums & Bass & Other & Avg & Params \\
         \midrule
        HDemucs & 8.13 & 8.24 & 8.76 & 5.59 & 7.68 & 42M \\
        BSRNN & 10.01 & 9.01 & 7.22 & 6.70 & 8.24 & 37M$\times$4 \\
        TFC-TDF V3 & 9.59 & 8.44 & 8.45 & 6.86 & 8.36 & 70M \\
        BS-RoFormer & 10.66 & 9.49 & \textbf{11.31} & 7.73 & 9.80 & 72M$\times$4 \\
        SCNet-L & 10.86 & \textbf{10.98} & 9.49 & 7.44 & 9.69 & 42M \\
        \midrule
        SCNet-S & 9.89 & 10.51 & 8.82 & 6.76 & 9.00 & 10M \\
        DTTNet & 10.12 & 7.74 & 7.45 & 6.92 & 8.06 & 5M$\times$4 \\
        \midrule
        Proposed & \textbf{10.92} & 10.93 & 10.04 & \textbf{7.95} & \textbf{9.96} & 5M$\times$4 \\
        Proposed-S & 9.87 & 9.13 & 8.81 & 6.79 & 8.65 & 2M$\times$4 \\
        \end{tabular}
    }
\centering
\end{table}

\subsection{Benchmark} \label{sec:benchmark}

In this section, we compare our proposed model to previous work using only the MUSDB-HQ dataset for training and evaluation. We selected seven most recent works with the best performance for comparison: HDemucs \cite{rouard2022hybrid}, BSRNN \cite{luo2023music}, TFC-TDF V3 \cite{kim2023sound}, BS-RoFormer \cite{lu2024music}, and SCNet-L \cite{tong2024scnet} as representatives of large models, alongside SCNet-S \cite{tong2024scnet} and DTTNet \cite{chen2024music} as representatives of lightweight models. The results are presented in Table \ref{tab:benchmark}. Our proposed method demonstrates on par or better performance compared to these previous works in the separation of \emph{vocals}, \emph{drums}, and \emph{other} musical stems. Despite having 13 times fewer parameters and significantly reduced training time and resource requirements, our method achieves an average SDR that is better than the state-of-the-art BS-RoFormer model. However, our model does not have an advantage in modeling \amy{\emph{bass} compared to BS-RoFormer. Higher bandsplit resolution might still be crucial in modeling instruments with lower frequency.}

% non-pitch instruments (\emph{drums}). In fact, models like TFC-TDF V3, DTTNet (utilizing TFC-TDF V3 blocks), and HDemucs exhibit lower performance on \emph{drums}. In contrast, models such as SCNet, BS-RoFormer, and BSRNN employ early layers (e.g., linear layers and convolution layers) on the frequency domain to capture relationships across frequency bins, tend to achieve better performance on \emph{drums}. Given that drum sounds span a broad frequency spectrum, such operations might be crucial for effective drum separation.

\begin{table}[]
\caption{Comparison between our proposed model and previous works when MoisesDB is used as an additional training data.}
\label{tab:benchmark_extra}
    \centering
    \resizebox{0.49\textwidth}{!}{
        \begin{tabular}{lccccccc}
         & Extra & Vocals & Drums & Bass & Other & Params \\
         \midrule
        HT Demucs & 800 & 9.20 & 10.08 & 10.39 & 6.32 & 42M \\
        BSRNN & 1750 & 10.47 & 10.15 & 8.16 & 7.08 & 37M$\times$4 \\
        BS-RoFormer & 450 & 12.72 & 12.91 & 13.32 & 9.01 & 93M$\times$4 \\
        SCNet-L & 235 & 11.11 & 11.23 & 9.86 & 7.51 & 42M \\
        \midrule
        Proposed & 235 & 11.10 & 11.10 & 10.77 & 8.28 & 5M$\times$4 \\
        \end{tabular}
    }
\centering

\end{table}

\subsection{Model Compression and Scalability} \label{sec:scalability}

To further optimize our model for lightweight inference, we reduced the convolution channels to $G=32$ and increased RoPE blocks to $N_{RoPE}=6$, resulting in 2.2 million parameters, comparable to SCNet-S with a 4-stem setup. As shown in Table \ref{tab:benchmark}, the proposed model, referred to as Proposed-S, achieves performance on par with SCNet-S for \emph{vocals}, \emph{bass}, and \emph{other} while utilizing slightly fewer parameters. However, it is inferior to SCNet-S on \emph{drums}. Moreover, our model experiences a greater performance decline with parameter reduction than SCNet. Nevertheless, the small proposed model still surpasses other works, such as DTTNet, TFC-TDF V3, and HDemucs, in average SDRs, despite having only 2.2 million parameters.

For lightweight models, it is often unclear how well they scale with larger-scale training data. To evaluate our proposed model with more training data, we integrated MoisesDB, with results detailed in Table \ref{tab:benchmark_extra}. Despite having only half the parameters, our model achieved better SDR scores compared to SCNet-L (which also used MoisesDB as additional data). BS-RoFormer achieved the highest SDRs when scaling up the training data. However, since their extended dataset is not available, a direct comparison with their results is not feasible.

When compared to HT Demucs, both BSRNN and SCNet have lower SDR on \emph{bass}, even though they perform better on other stems. Interestingly, BS-RoFormer also shows the highest performance on \emph{bass} relative to other instruments. This suggests that when scaling up the training data, \amy{transformer architecture might be a better choice for \emph{bass}}. Our proposed model, which utilizes transformer architecture for sequence modeling in the bottleneck, has achieved superior results on \emph{bass} while having a smaller number of parameters and less training data compared to HT Demucs.

\section{Conclusion}

In this work, we propose a lightweight model \amy{to show that without blindly scaling up the model's parameters, a small model with 13x fewer parameters than BS-RoFormer can achieve comparable performance on MUSDB-HQ benchmark}. Drawing inspiration from other works, we boost the average SDR by 1.9 dB. Through step-by-step ablation studies, we analyze the impact of different components on performance across various instruments. Our model also scales well, outperforming SCNet-L with half the parameters. However, we identify limitations in more lightweight scenarios, notably a drop in performance for \emph{drums} and lower SDR scores compared to SCNet-S.

In the future, we plan to conduct a comprehensive study focusing on improving \emph{drum} and \emph{bass} separation performance. Additionally, we aim to evaluate the model's separation capabilities on instruments beyond the standard four stems. \amy{Although not reflected on the SDRs, the proposed band splitting sometimes introduces perceivable artifacts. We also plan to investigate and improve the perceptual performance in the future.}

% -------------------------------------------------------------------------
% Either list references using the bibliography style file IEEEtran.bst

\clearpage
% The \IEEEtriggeratref{XX} command can be used to move to the next column before the XX-th reference
% to balance the two columns of the reference section
% \IEEEtriggeratref{XX}
\bibliographystyle{IEEEtran}
\bibliography{refs25}

\begin{thebibliography}{10}
\providecommand{\url}[1]{#1}
\csname url@samestyle\endcsname
\providecommand{\newblock}{\relax}
\providecommand{\bibinfo}[2]{#2}
\providecommand{\BIBentrySTDinterwordspacing}{\spaceskip=0pt\relax}
\providecommand{\BIBentryALTinterwordstretchfactor}{4}
\providecommand{\BIBentryALTinterwordspacing}{\spaceskip=\fontdimen2\font plus
\BIBentryALTinterwordstretchfactor\fontdimen3\font minus \fontdimen4\font\relax}
\providecommand{\BIBforeignlanguage}[2]{{%
\expandafter\ifx\csname l@#1\endcsname\relax
\typeout{** WARNING: IEEEtran.bst: No hyphenation pattern has been}%
\typeout{** loaded for the language `#1'. Using the pattern for}%
\typeout{** the default language instead.}%
\else
\language=\csname l@#1\endcsname
\fi
#2}}
\providecommand{\BIBdecl}{\relax}
\BIBdecl

\bibitem{veire2018raw}
L.~V. Veire and T.~De~Bie, ``From raw audio to a seamless mix: creating an automated dj system for drum and bass,'' \emph{EURASIP Journal on Audio, Speech, and Music Processing}, vol. 2018, no.~1, p.~13, 2018.

\bibitem{rafii2018overview}
Z.~Rafii, A.~Liutkus, F.-R. St{\"o}ter, S.~I. Mimilakis, D.~FitzGerald, and B.~Pardo, ``An overview of lead and accompaniment separation in music,'' \emph{IEEE Trans. Audio, Speech, Lang. Process.}, vol.~26, no.~8, pp. 1307--1335, 2018.

\bibitem{cifka-2024-jam-alt}
O.~C\'ifka, H.~Schreiber, L.~Miner, and F.-R. St\"oter, ``Lyrics transcription for humans: A readability-aware benchmark,'' in \emph{Proc. ISMIR}.\hskip 1em plus 0.5em minus 0.4em\relax ISMIR, 2024.

\bibitem{nakano2019joint}
T.~Nakano, K.~Yoshii, Y.~Wu, R.~Nishikimi, K.~W.~E. Lin, and M.~Goto, ``Joint singing pitch estimation and voice separation based on a neural harmonic structure renderer,'' in \emph{Proc. WASPAA}.\hskip 1em plus 0.5em minus 0.4em\relax IEEE, 2019, pp. 160--164.

\bibitem{donahue2023singsong}
C.~Donahue, A.~Caillon, A.~Roberts, E.~Manilow, P.~Esling, A.~Agostinelli, M.~Verzetti, I.~Simon, O.~Pietquin, N.~Zeghidour \emph{et~al.}, ``Singsong: Generating musical accompaniments from singing,'' \emph{arXiv preprint arXiv:2301.12662}, 2023.

\bibitem{wang2024mel}
J.-C. Wang, W.-T. Lu, and J.~Chen, ``Mel-roformer for vocal separation and vocal melody transcription,'' in \emph{Proc. ISMIR}, 2024.

\bibitem{hennequin2020spleeter}
R.~Hennequin, A.~Khlif, F.~Voituret, and M.~Moussallam, ``Spleeter: a fast and efficient music source separation tool with pre-trained models,'' \emph{Journal of Open Source Software}, vol.~5, no.~50, p. 2154, 2020.

\bibitem{stoter2019open}
F.-R. St{\"o}ter, S.~Uhlich, A.~Liutkus, and Y.~Mitsufuji, ``Open-unmix-a reference implementation for music source separation,'' \emph{Journal of Open Source Software}, vol.~4, no.~41, p. 1667, 2019.

\bibitem{takahashi2020d3net}
N.~Takahashi and Y.~Mitsufuji, ``D3net: Densely connected multidilated densenet for music source separation,'' \emph{arXiv preprint arXiv:2010.01733}, 2020.

\bibitem{chandna2017monoaural}
P.~Chandna, M.~Miron, J.~Janer, and E.~G{\'o}mez, ``Monoaural audio source separation using deep convolutional neural networks,'' in \emph{Latent Variable Analysis and Signal Separation: 13th International Conference}.\hskip 1em plus 0.5em minus 0.4em\relax Springer, 2017, pp. 258--266.

\bibitem{Jansson2017SingingVS}
A.~Jansson, E.~J. Humphrey, N.~Montecchio, R.~M. Bittner, A.~Kumar, and T.~Weyde, ``Singing voice separation with deep u-net convolutional networks,'' in \emph{Proc. ISMIR}, 2017.

\bibitem{rouard2022hybrid}
S.~Rouard, F.~Massa, and A.~D{\'e}fossez, ``Hybrid transformers for music source separation,'' in \emph{Proc. ICASSP}, 2023.

\bibitem{luo2023music}
Y.~Luo and J.~Yu, ``Music source separation with band-split rnn,'' \emph{IEEE Trans. Audio, Speech, Lang. Process.}, vol.~31, pp. 1893--1901, 2023.

\bibitem{kim2023sound}
M.~Kim, J.~H. Lee, and S.~Jung, ``Sound demixing challenge 2023 music demixing track technical report: Tfc-tdf-unet v3,'' \emph{arXiv preprint arXiv:2306.09382}, 2023.

\bibitem{ronneberger2015u}
O.~Ronneberger, P.~Fischer, and T.~Brox, ``U-net: Convolutional networks for biomedical image segmentation,'' in \emph{Medical image computing and computer-assisted intervention}.\hskip 1em plus 0.5em minus 0.4em\relax Springer, 2015, pp. 234--241.

\bibitem{lu2024music}
W.-T. Lu, J.-C. Wang, Q.~Kong, and Y.-N. Hung, ``Music source separation with band-split rope transformer,'' in \emph{Proc. ICASSP}, 2024, pp. 481--485.

\bibitem{chen2024music}
J.~Chen, S.~Vekkot, and P.~Shukla, ``Music source separation based on a lightweight deep learning framework (dttnet: Dual-path tfc-tdf unet),'' in \emph{Proc. ICASSP}.\hskip 1em plus 0.5em minus 0.4em\relax IEEE, 2024, pp. 656--660.

\bibitem{conformer}
A.~Gulati, J.~Qin, C.-C. Chiu, N.~Parmar, Y.~Zhang, J.~Yu, W.~Han, S.~Wang, Z.~Zhang, Y.~Wu, and R.~Pang, ``Conformer: Convolution-augmented transformer for speech recognition.'' in \emph{Proc. Interspeech}.\hskip 1em plus 0.5em minus 0.4em\relax ISCA, 2020, pp. 5036--5040.

\bibitem{della2024resource}
L.~Della~Libera, C.~Subakan, M.~Ravanelli, S.~Cornell, F.~Lepoutre, and F.~Grondin, ``Resource-efficient separation transformer,'' in \emph{Proc. ICASSP}.\hskip 1em plus 0.5em minus 0.4em\relax IEEE, 2024, pp. 761--765.

\bibitem{subakan2021attention}
C.~Subakan, M.~Ravanelli, S.~Cornell, M.~Bronzi, and J.~Zhong, ``Attention is all you need in speech separation,'' in \emph{Proc. ICASSP}.\hskip 1em plus 0.5em minus 0.4em\relax IEEE, 2021, pp. 21--25.

\bibitem{yip2024spgm}
J.~Q. Yip, S.~Zhao, Y.~Ma, C.~Ni, C.~Zhang, H.~Wang, T.~H. Nguyen, K.~Zhou, D.~Ng, E.~S. Chng \emph{et~al.}, ``Spgm: Prioritizing local features for enhanced speech separation performance,'' in \emph{Proc. ICASSP}, 2024, pp. 326--330.

\bibitem{Yang2024FspenAU}
L.~Yang, W.~Liu, R.~Meng, G.~Lee, S.~Baek, and H.-G. Moon, ``Fspen: an ultra-lightweight network for real time speech enahncment,'' \emph{Proc. ICASSP}, pp. 10\,671--10\,675, 2024.

\bibitem{luo2021ultra}
Y.~Luo, C.~Han, and N.~Mesgarani, ``Ultra-lightweight speech separation via group communication,'' in \emph{Proc. ICASSP}.\hskip 1em plus 0.5em minus 0.4em\relax IEEE, 2021, pp. 16--20.

\bibitem{tong2024scnet}
W.~Tong, J.~Zhu, J.~Chen, S.~Kang, T.~Jiang, Y.~Li, Z.~Wu, and H.~Meng, ``Scnet: Sparse compression network for music source separation,'' in \emph{Proc. ICASSP}, 2024, pp. 1276--1280.

\bibitem{venkatesh2024real}
S.~Venkatesh, A.~Benilov, P.~Coleman, and F.~Roskam, ``Real-time low-latency music source separation using hybrid spectrogram-tasnet,'' in \emph{Proc. ICASSP}.

\bibitem{kim23i_interspeech}
J.~Kim and H.-G. Kang, ``Contrastive learning based deep latent masking for music source separation,'' in \emph{Proc. Interspeech}, 2023, pp. 3709--3713.

\bibitem{StollerED18}
D.~Stoller, S.~Ewert, and S.~Dixon, ``Wave-u-net: {A} multi-scale neural network for end-to-end audio source separation,'' in \emph{Proc. ISMIR}, 2018, pp. 334--340.

\bibitem{guso2022loss}
E.~Gus{\'o}, J.~Pons, S.~Pascual, and J.~Serr{\`a}, ``On loss functions and evaluation metrics for music source separation,'' in \emph{Proc. ICASSP}, 2022, pp. 306--310.

\bibitem{Dfossez2019DemucsDE}
A.~D{\'e}fossez, N.~Usunier, L.~Bottou, and F.~R. Bach, ``Demucs: Deep extractor for music sources with extra unlabeled data remixed,'' \emph{ArXiv}, vol. abs/1909.01174, 2019.

\bibitem{SuJF20-0}
J.~Su, Z.~Jin, and A.~Finkelstein, ``Hifi-gan: High-fidelity denoising and dereverberation based on speech deep features in adversarial networks,'' in \emph{Proc. Interspeech}.\hskip 1em plus 0.5em minus 0.4em\relax ISCA, 2020, pp. 4506--4510.

\bibitem{yamamoto2020parallel}
R.~Yamamoto, E.~Song, and J.-M. Kim, ``Parallel {WaveGAN}: {A} fast waveform generation model based on generative adversarial networks with multi-resolution spectrogram,'' in \emph{Proc. ICASSP}, 2020, pp. 6199--6203.

\bibitem{musdb18-hq}
\BIBentryALTinterwordspacing
Z.~Rafii, A.~Liutkus, F.-R. Stöter, S.~I. Mimilakis, and R.~Bittner, ``Musdb18-hq - an uncompressed version of musdb18,'' Aug. 2019. [Online]. Available: \url{https://doi.org/10.5281/zenodo.3338373}
\BIBentrySTDinterwordspacing

\bibitem{moisesdb}
I.~Pereira, F.~Ara{\'{u}}jo, F.~Korzeniowski, and R.~Vogl, ``Moisesdb: {A} dataset for source separation beyond 4-stems,'' in \emph{Proc. ISMIR}, 2023, pp. 619--626.

\bibitem{stoter20182018}
F.-R. St{\"o}ter, A.~Liutkus, and N.~Ito, ``The 2018 signal separation evaluation campaign,'' in \emph{Latent Variable Analysis and Signal Separation: 14th International Conference}.\hskip 1em plus 0.5em minus 0.4em\relax Springer, 2018, pp. 293--305.

\end{thebibliography}
% or list them by yourself:

% \bibitem{waspaaweb}
% {WASPAA Website}, \url{http://www.waspaa.com}.

% \bibitem{IEEEXploreReqs}
% {IEEE {X}plore {R}equirements}, \url{https://conferences.ieeeauthorcenter.ieee.org/write-your-paper/meet-ieee-xplore-requirements/}.

% \bibitem{eWilliams1999}
% E.~Williams, \emph{Fourier Acoustics: Sound Radiation and Nearfield Acoustic Holography}.\hskip 1em plus 0.5em minus 0.4em\relax London, UK: Academic Press, 1999.

% \bibitem{cJones2003}
% C.~Jones, A.~Smith, and E.~Roberts, ``A sample paper in conference proceedings,'' in \emph{Proc. ICASSP}, vol.~II, Apr. 2003, pp. 803--806.

% \bibitem{aSmith2000}
% A.~Smith, C.~Jones, and E.~Roberts, ``A sample paper in journals,'' \emph{IEEE Trans. Signal Process.}, vol.~62, pp. 291--294, Jan. 2000.

% \end{thebibliography}

\end{document}